\input{psfig}
\documentstyle[11pt,newpasp,twoside]{article}
\def\edcomment#1{\iffalse\marginpar{\raggedright\sl#1\/}\else\relax\fi}
\reversemarginpar

\begin{document}
\title{Planetary migration in protoplanetary disks}
  \author{A. Del Popolo}
\affil{Dipartimento di Matematica, Universit\`{a} Statale di Bergamo,
Piazza Rosate, 2 - I 24129 Bergamo, ITALY}
\affil{Feza G\"ursey Institute, P.O. Box 6 \c Cengelk\"oy, Istanbul,
     Turkey}

\begin{abstract}

In the current paper, we further develop the model for the migration of
planets introduced in Del Popolo et al. (2001) and extended to time-dependent planetesimal 
accretion discs in Del Popolo \& Ek\c{s}i (2002). 
We use a method developed by Stepinski \& Valageas (1996, 1997), that is
able to simultaneously follow the evolution of gas and solid particles
for up to $10^7 {\rm yr}$. The disc model is coupled to 
the migration model introduced in Del Popolo et al. (2001) in order to 
obtain the migration rate of the planet
in the planetesimal disc. We find that in the case of discs 
having total mass
of $10^{-3}-0.1 M_{\odot}$, and $0.1<\alpha<0.0001$, planets can migrate inward a large distance
while if $M<10^{-3} M_{\odot}$ the planets remain almost in their initial 
position for $0.1<\alpha<0.01$ and only in the case $\alpha<0.001$ 
the planets move to a minimum value of orbital radius of $\simeq 2 {\rm AU}$.
The model gives a good description of the observed distribution of planets 
in the period range 0-20 days.

\end{abstract}

\section{Introduction}

In order to study the formation of planetary systems it is
necessary to study the global evolution of solid material which
constitutes, together with gas, the protoplanetary discs. 
In the following, we introduce a time-dependent accretion disc
model that shall be used in the next sections to study planets
migration.

It is usually assumed that the surface density in planetesimals is proportional to 
that of gas (Lin \& Papaloizou (1980, page 47); Murray et al. (1998))
and that the spatial structure and the time evolution of the
surface density, $\Sigma_{\rm s}(R,t)$, is such that $\Sigma_{\rm s}(R,t) \propto \Sigma(R,t)$.
Although this approximation has been and is widely used in literature, a better way to proceed is to 
directly calculate the evolution of solids in the disc.  
We know that the evolution of the surface density of gas is described 
by a diffusive-type equation, while that of solid particles is an advection-diffusion equation:
\begin{equation}
\frac{\partial \Sigma_{\rm s} }{\partial t}=\frac{3}{r}\frac{\partial }{\partial r}%
\left[ r^{1/2}\frac{\partial }{\partial r}(\nu_{\rm s} \Sigma_{\rm s}
r^{1/2})\right]+
\frac{1}{r}\frac{\partial }{\partial r}\left[ \frac{2r
\Sigma _{s}\langle \overline{v}_{\phi }\rangle _{s}}{\Omega _{k}t_{s}}\right]  
\label{advdiffusee}
\end{equation}
where $\nu_{\rm s}=\frac{\nu}{{\rm Sc}}$, the Schmidt number, Sc, is given by:
\begin{equation}
{\rm Sc}=\left( 1+\Omega _{\rm k}t_{\rm s}\right) \sqrt{1+\frac{\overline{\bf v}^2}{V_{\rm t}^2}}
\end{equation}
where ${\bf v}$ is the relative velocity between a particle and the gas, $V_{\rm t}$ the 
turbulent velocity, $\Omega _{\rm k}$ is the Keplerian angular velocity, $t_{\rm s}$ the so called stopping time.

Stepinski \& Valageas (1996, 1997) developed a method that, using a series of simplifying assumptions, is able to simultaneously follow the evolution of gas and solid particles due to gas-solid-coupling, coagulation, sedimentation, and evaporation/condensation for up to $10^7 {\rm yr}$. 
%
%

The equation of gas evolution  
is solved by means of an implicit scheme and the evolution of gas
is computed independently from the evolution of particles. 
At every time step the quantities
needed for evaluating the change in the mass distribution
of solids are calculated and the change itself is computed
from Eq. (1) using the operator splitting method. In
such a method the advective term in Eq. (1) is treated by the
numerical method of characteristics, whereas an implicit
scheme is applied to the diffusion term. The obtained 
distribution of solid material is then modifed because of the
existence of the evaporation radius, and the mass distribu-
tion of the vapor is calculated using the implicit scheme.
Finally, the new particle size distribution is calculated be-
fore proceeding to the next time step.

We suppose that a single planet moves in a
planetesimal disc under the influence of the gravitational force
of the Sun. The equation of motion of the planet can be written as:
\begin{equation}
{\bf \ddot r}= {\bf F}_{\odot}
+{\bf R}
\end{equation}
(Melita \& Woolfson 1996), where the term ${\bf F}_{\odot}$ represents
the force per unit mass from the Sun, while ${\bf R}$ is the dissipative
force (the dynamical friction term-see Melita \& Woolfson 1996). 

In order to take into account dynamical friction, we need a 
suitable formula for a disc-like structure such as the 
protoplanetary disc. 

\indent We assume that the matter-distribution is disc-shaped, then 
we have that:
\begin{equation}
{\bf R}=-k_{\parallel}v_{1 \parallel} {\bf e_{\parallel}}-
k_{\perp}v_{1 \perp}{\bf e_{\perp}}
\label{eq:dyn}
\end{equation}
where ${\bf e_{\parallel}}$ and ${\bf e_{\perp}}$ are two versors
parallel and perpendicular to the disc plane and 
$k_{\parallel}$ and $k_{\perp}$ are given in Del Popolo \& Ek\c{s}i (2002).

\section{Results}

\indent Our model starts with a fully formed gaseous giant planet of
$1 M_{\rm J}$ at $5.2$ AU.
According to several evidences showing that the disc lifetimes range 
from
$ 10^5$ yr to $10^7$ ~yr (Strom et al. 1993; Ruden \& Pollack 1991), 
we assume that the disc has a nominal effective lifetime of $10^6$ 
years (Zuckerman et al. 1995). 
We integrated the model introduced in the previous section for
several values of disc
masses: $M_{\rm D}=0.1$, 0.01, 0.001, 0.0001 $M_{\odot}$, and several values of $\alpha$.\\

The results of the disc model are plotted in Fig.1.
Fig. 1a shows the evolution of the density of the gas component of the disc in the 
case $M=0.1 M_{\odot}$ and $\alpha=0.1$ (see figure).
%
%
As shown in Fig. 1b, in agreement with Stepinski \& Valageas (1997) and Kornet et al. (2001),
the most important result of the low-mass models 
calculation (discs with $M<0.1 M_{\odot}$, and angular momentum $\simeq 10^{52} {\rm g cm^2/s}$) 
is that such models lead to the survival of solid
material, as can be seen from the evolution of the particle size,
or by the emergence of the converged, nonvanishing surface density distribution 
of solids. 

In the successive figures (2a-2d), we plot the evolution of
semi-major axis of the planet (see figure). 
Summarizing figures 2a-2d, according to the final distribution of planets distances, 
the present model predicts that, unless the disc mass is very small $M \simeq 0.0001 M_{\odot}$,
planets tend to move close to the central star to distances of the order of $0.03 {\rm AU}$.
It is also evident that it is possible to find a 
planet at any distance from their locations of
formation and very small distances from the parent star for peculiar values 
of the parameters $\alpha$ and $M$. 
 
In Fig. 3a, in order to show the predictions of the model for what concerns the distribution 
of planets in the inner part of the disc, we plot the fraction of planets in the orbital period 
range 0-20 days calculated using the model of this paper. Fig. 3b, represents the same distribution 
obtained with the data given in www.exoplanets.org (see also Kuchner \& Lecar 2002).

According to Kuchner \& Lecar 2002, disc temperature determines the orbital radii of 
the innermost surviving planets, similarly to our model. 
As streseed in DP1, DP2, 
the model has not the drawback of Murray et al. 1998 model, namely that of requiring a too large 
disc mass for migration and at the same time has the advantage of Murray et al. 1998 model of 
having an intrinsic natural mechanism that provides halting of migration.

\begin{figure}
\label{Fig. 1} \centerline{\hbox{(1a)
\psfig{figure=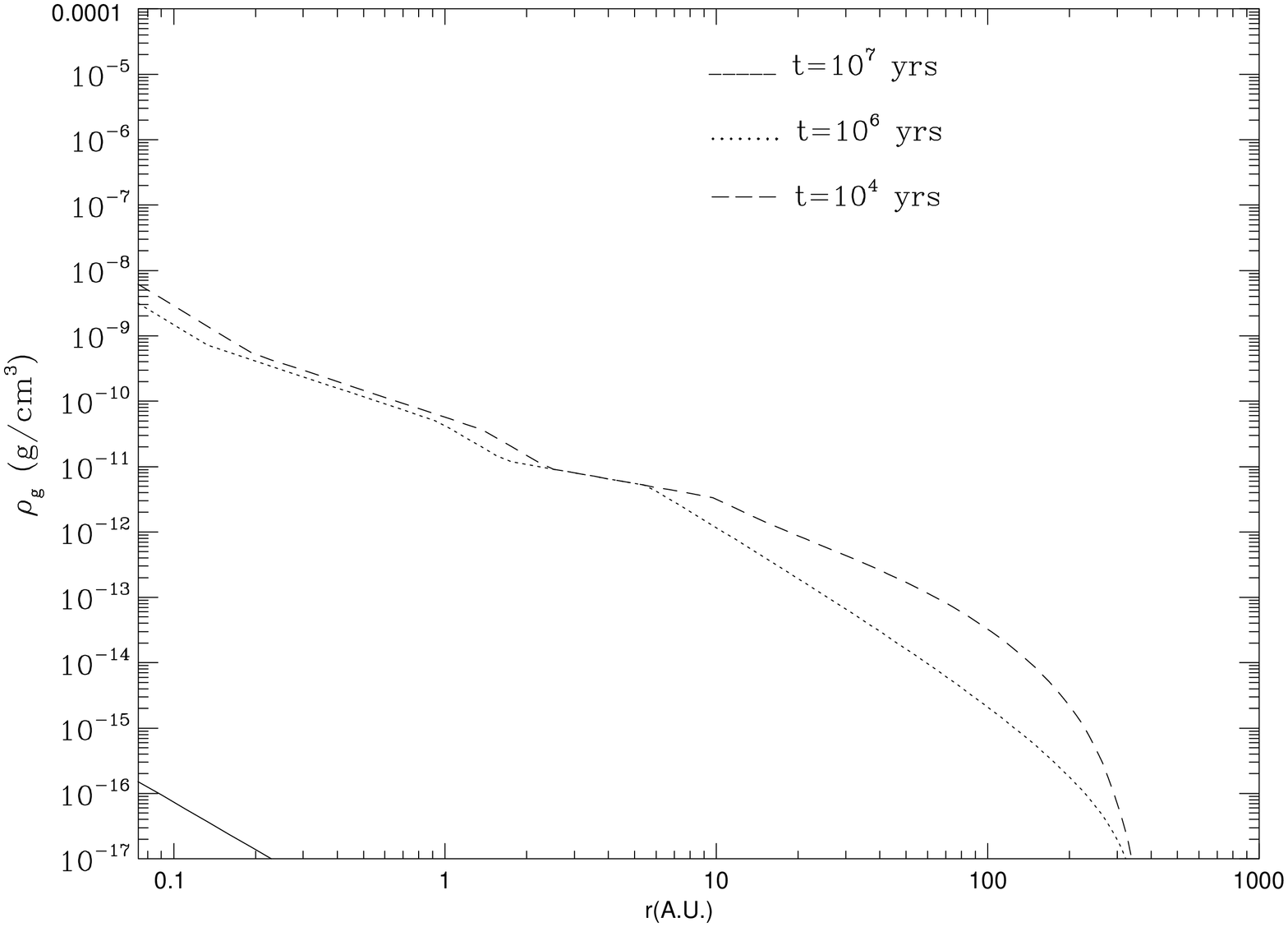,width=4.2cm,angle=270} (1b)
\psfig{figure=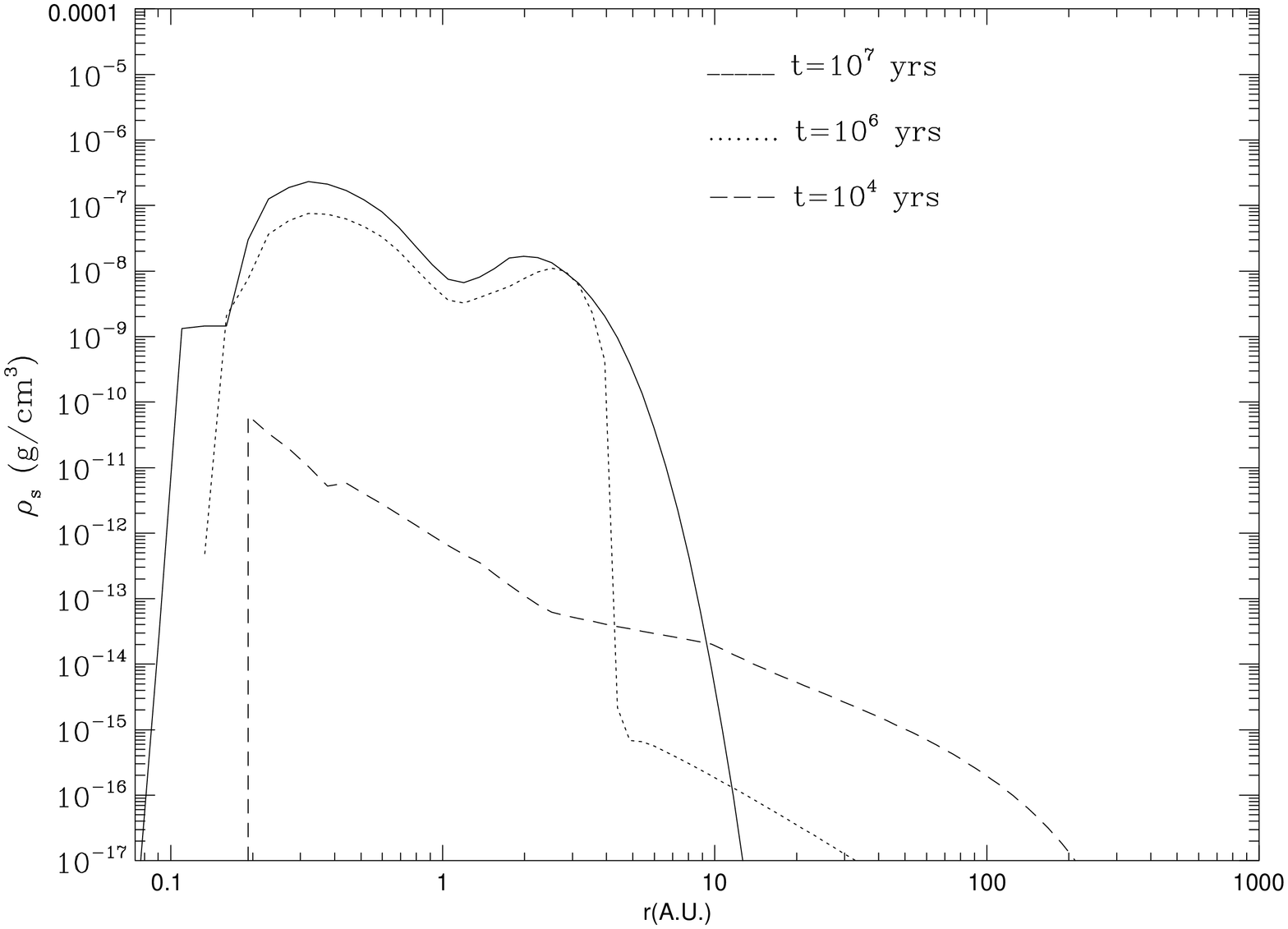,width=4.2cm,angle=270}
}}
\label{Fig. 1} \centerline{\hbox{(2a)
\psfig{figure=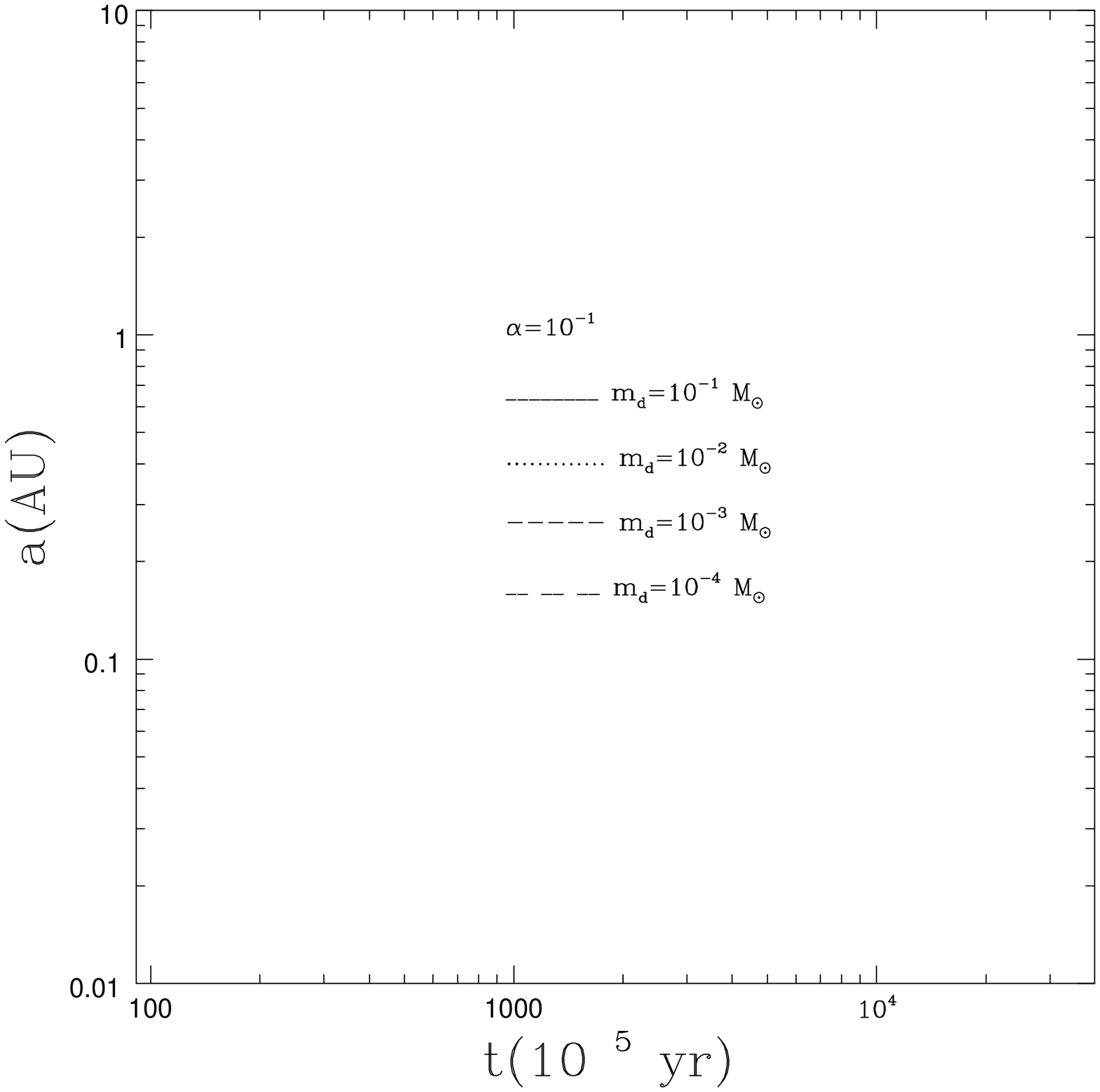,width=5cm} (2b)
\psfig{figure=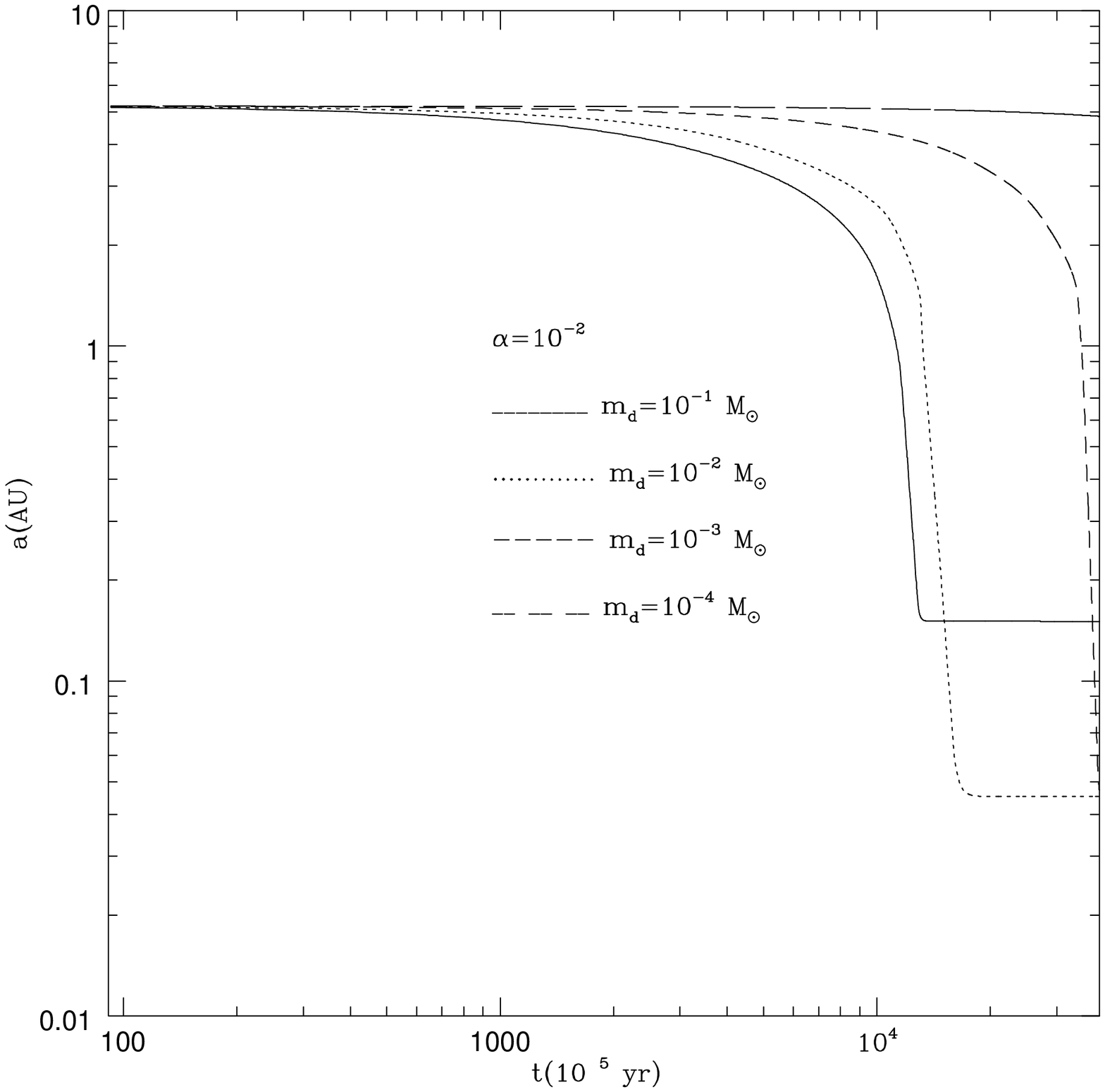,width=5cm}
}}
\label{Fig. 1} \centerline{\hbox{(2c)
\psfig{figure=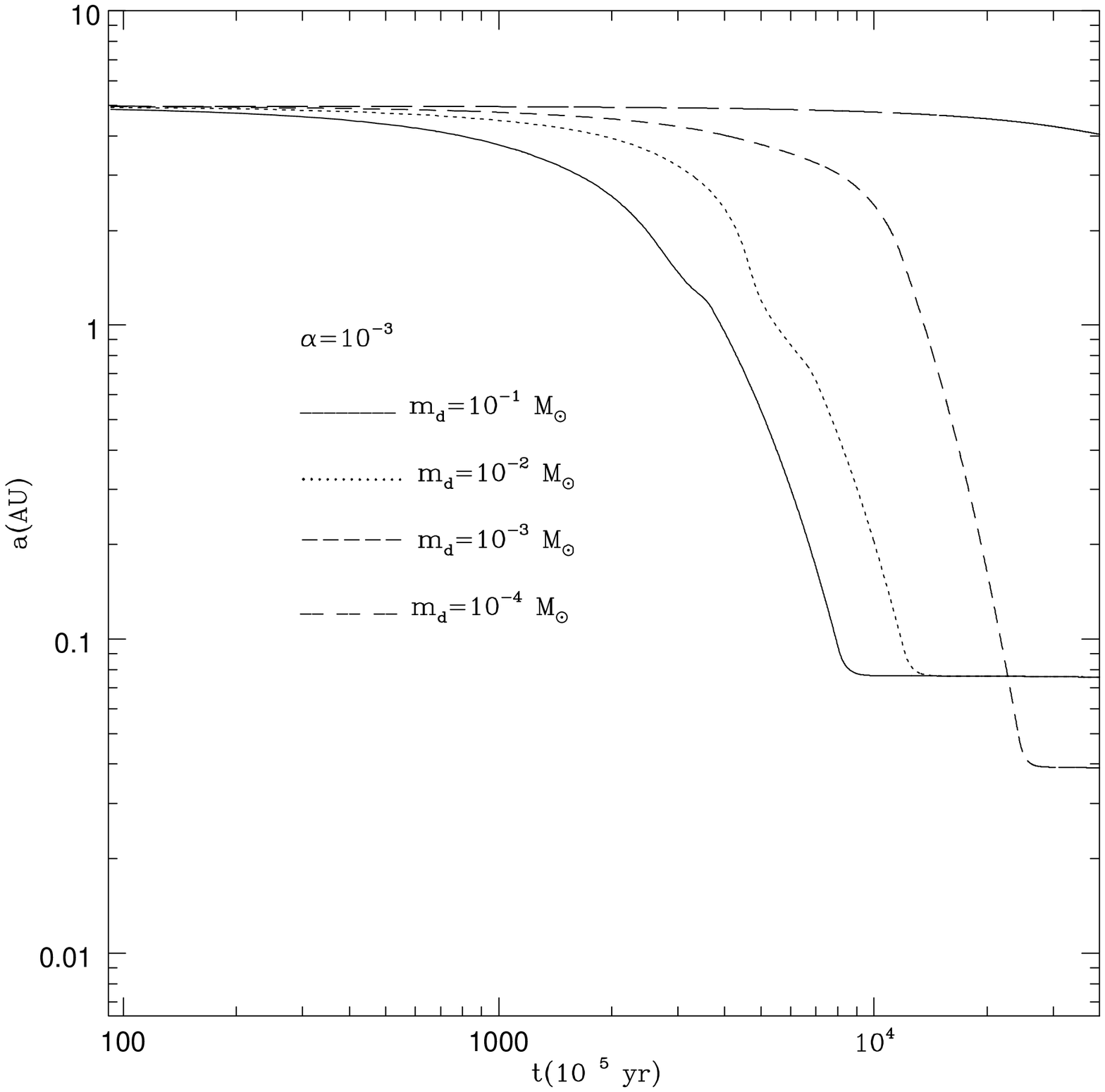,width=5cm} (2d)
\psfig{figure=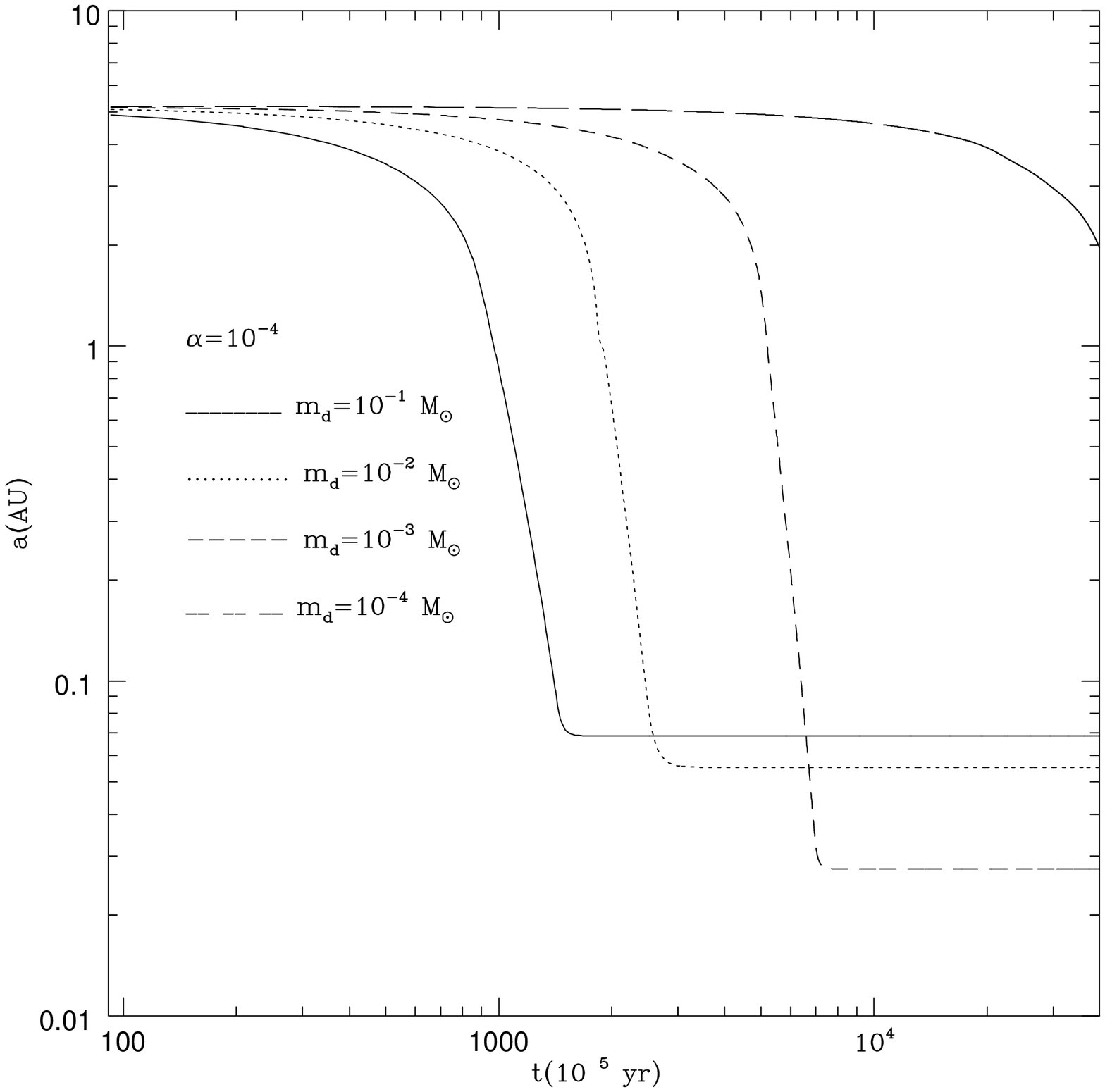,width=5cm}
}}
\label{Fig. 1} \centerline{\hbox{(3a)
\psfig{figure=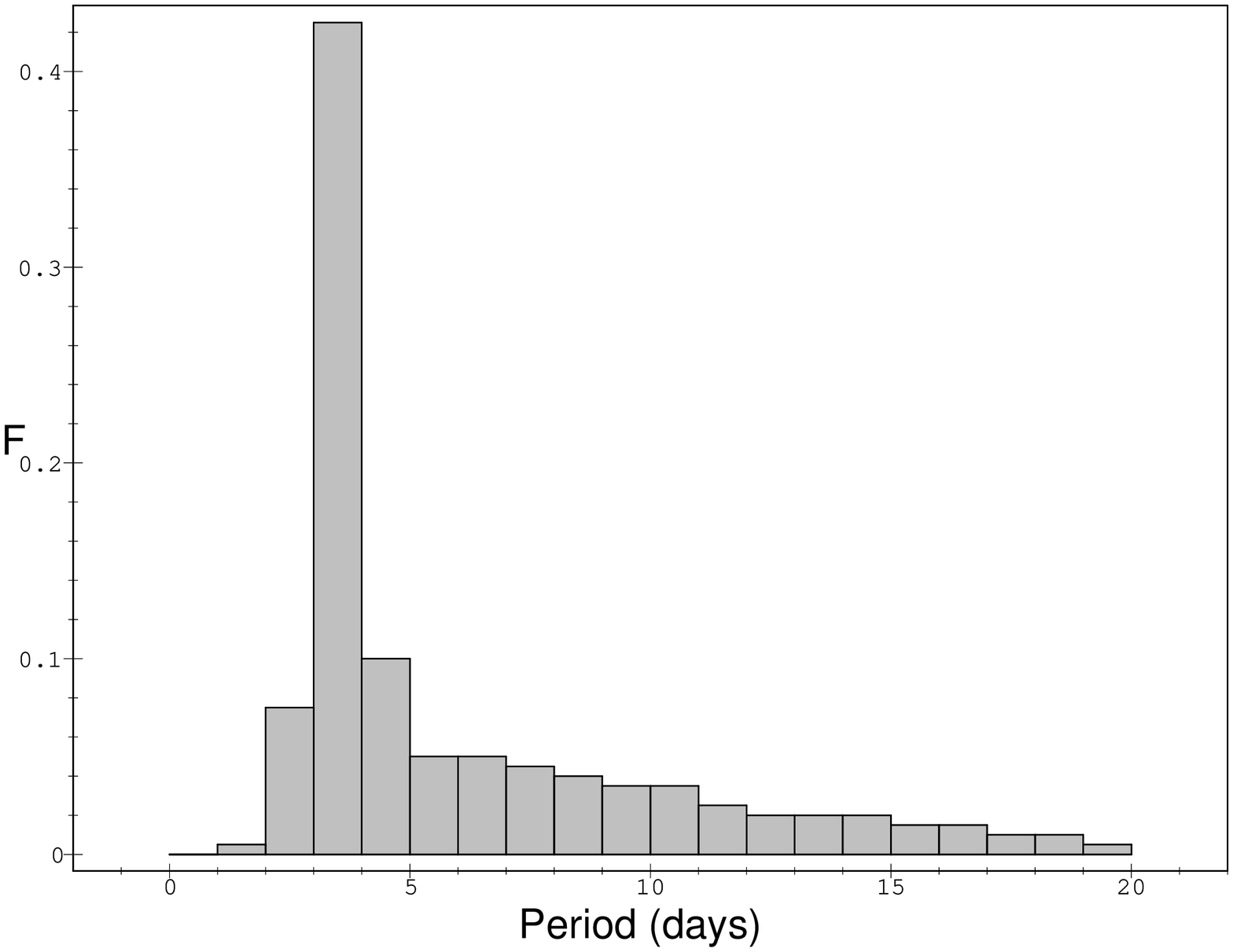,width=4.6cm,height=4.6cm,angle=270} (3b)
\psfig{figure=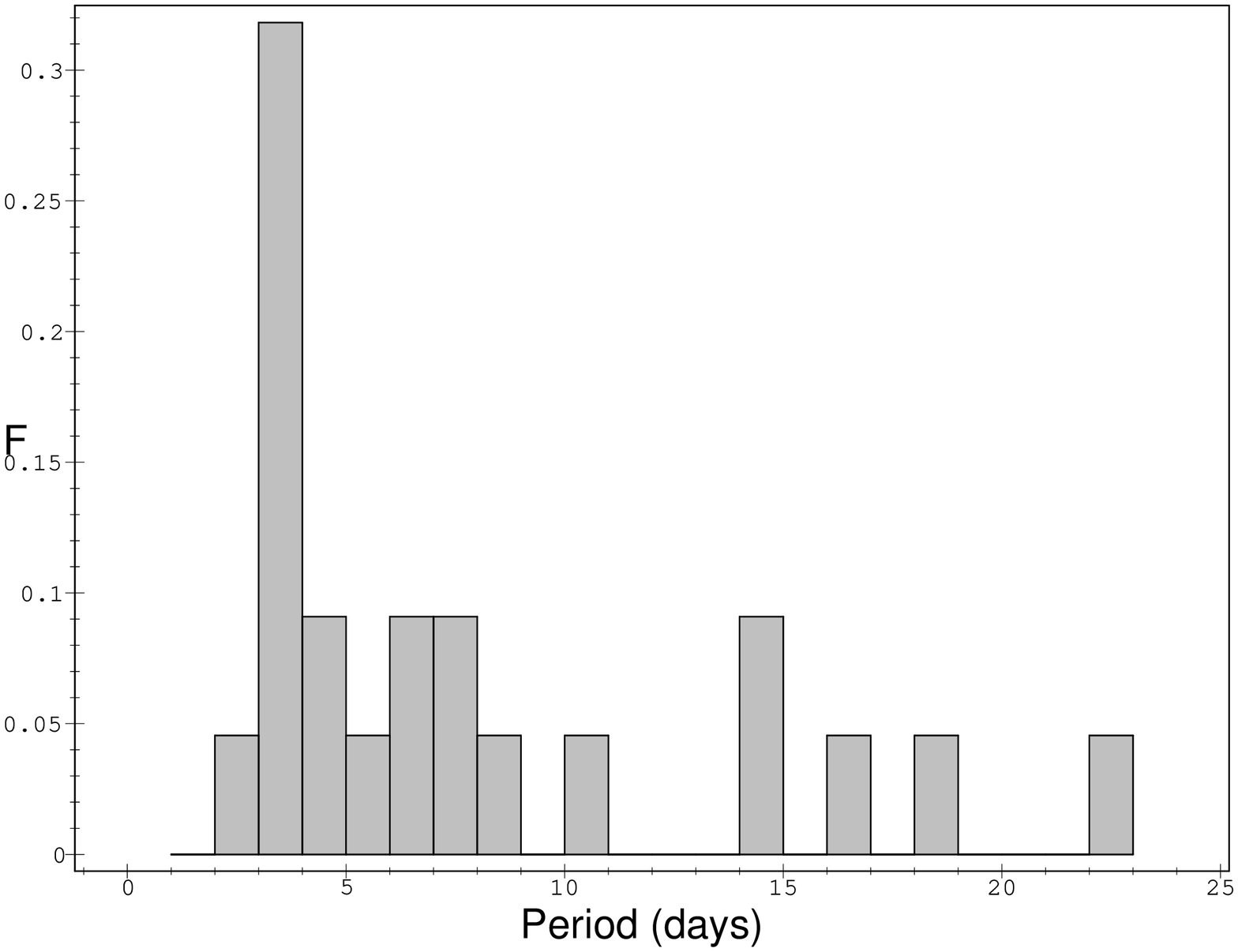,width=4.6cm,height=4.6cm,angle=270}
}}
\caption[]{(a) Evolution of gas for a disc with
$M_d=0.1 M_{\odot}$ and $\alpha=0.1$ at $t=10^4 {\rm yrs}$ (dashed line),
$t=10^6 {\rm yrs}$ (dotted line) and $t=10^7 {\rm yrs}$ (solid line). 
(b) Same as Fig. 1a but for solids.
}
\caption[]{(a) The evolution of the a(t) of a Jupiter-mass planet,
$M=1 M_{\rm J}$ in a planetesimal disc 
for $\alpha=0.1$ and several values of $M_{\rm d}$,
0.1  (solid line), 0.01 (dotted line),
0.001 $M_{\odot}$ (short-dashed line) and 0.0001
(long-dashed line). (b,c, d) Same as Fig. (6a) but with $\alpha=0.01, 0.001, 0.0001$
}
\caption[]{(a) Fraction of planets having orbital periods in the range 0-20 days, 
calculated according the model of this paper. 
(b) Fraction of planets having orbital periods in the range 0-20 days, 
calculated according to data (see www.exoplanets.org).}
\end{figure}
\end{document}